\documentstyle[aps,multicol,psfig]{revtex}

\begin{document}
\draft

\title{Two pairing parameters in superconducting grains}

\author{M. Schechter$^1$, J. von Delft$^2$, Y. Imry$^3$, and Y. Levinson$^3$}

\address{$^1$ The Racah Institute of Physics, Hebrew University, Jerusalem 
91904, Israel 
\newline
	$^2$ Sektion Physik and Center for NanoScience,
Ludwig-Maximilians-Universit\"at M\"unchen, Theresienstr. 37, D-80333,
Germany 
\newline
	$^3$Department of Condensed Matter Physics, The 
Weizmann Institute of Science, Rehovot 76100, Israel }

%\date{Oct 13, 2002}

\maketitle

\vspace{0.4cm}

\begin{abstract}
Unlike bulk superconductivity, where one energy scale, the energy gap,
characterizes pairing correlations, we show that in small
superconducting grains there exist two different such quantities. The
first characterizes cumulative properties of the grain, such as the
condensation energy, and the second single-particle properties. To
describe these two energy scales, we define two corresponding pairing
parameters, and show that although both reduce to the bulk gap for
large grains, this occurs at different size scales.

\end{abstract}

\vspace{0.5cm}

\begin{multicols}{2}

\section{Introduction}

The question of how superconductivity is affected by small sample size, 
raised already by Anderson in 1959\cite{And59}, has
experienced a recent revival of interest, which started with the
experimental work of Ralph, Black and Tinkham\cite{BRT96,RBT97}. In
these experiments it was shown that small superconducting grains, with
size much smaller than the coherence length, but level spacing $d$
smaller than the bulk gap $\Delta$, have a gap of order $2
\Delta$ in their tunneling spectrum, and grains in the regime $d
\gtrsim \Delta$ (to be called ``ultrasmall'' regime) do not. 
Since the pairing
parameter is the basic quantity in bulk superconductivity, 
several efforts have been made to define pairing
parameters which are relevant in the regime of small
grains\cite{DZGT96,ML97,BD98,BD99,RCR99,DR01}, and reduce to $\Delta$ in 
the bulk limit.
The need for such new definitions arises for two apparent reasons: (i)
for ultrasmall grains, in which $d > \Delta$, the quantity
$\Delta$ has no direct physical meaning; 
(ii) in both ultrasmall grains, and small grains for which $d \lesssim
\Delta$, if the grains are isolated the appropriate ensemble is the 
canonical one, in which the
usual definition of $\Delta$ that is used in the grand canonical
ensemble [see $\Delta_{\rm g.c.}$ in Eq.~(\ref{deltagc})] is
trivially zero.  In this paper we show that there is a third,
fundamental reason for the inadequacy of $\Delta_{\rm g.c.}$ to
describe pairing correlations in small superconducting grains.  Unlike
the situation in bulk superconductors, in which all the various
superconducting properties can be characterized by one energy scale,
$\Delta$, in general there exist two distinct energy scales that
characterize pairing correlations, and the difference between them
becomes important particularly for small grains. The first such 
energy scale, which we denote by $\Delta_{\rm s.p.}$, characterizes
single-particle properties, such as excitation energies and even-odd
effects. The second, denoted as $\Delta_{\rm cu}$, characterizes
cumulative properties, such as the condensation energy [see
Eqs.~(\ref{condensationdelta}),~(\ref{continuouslambda})], to which
pairing correlations of all the levels up to $\omega_{\rm D}$
contribute.  The reason why these two scales are in general distinct
is that levels far from the Fermi energy, namely those with energy
$|\varepsilon - E_{\rm F}|$ between $\Delta$ and the Debye frequency
$\omega_{\rm D}$ (to be called ``not condensed'' or ``far''
  levels), make a more significant contribution to cumulative
properties than to single-particle ones.  (For a discussion on the
role of the far levels in superconducting grains and persistent
currents in normal metal rings see Refs.~\onlinecite{SILD01,SOIL02}).
The contribution of the far levels to physical properties turns out to
be proportional to the level spacing; in the large grain limit where
the level spacing becomes exceedingly small, this contribution can
thus be neglected. In this limit, the single-particle and cumulative
properties are therefore both determined by the correlations of
only the ``condensed levels'', i.e.\ those within $\Delta$ of 
$E_{\rm F}$, so that the two scales $\Delta_{\rm cu}$ and 
$\Delta_{\rm s.p.}$ become identical.

Previous attempts\cite{DZGT96,BD98,BD99,RCR99,DR01} to define, in terms 
of pair correlation functions, a pairing
parameter adequate to describe small grains in the canonical ensemble
resulted in parameters characterizing cumulative properties of the
grain, but not single particle ones. 
We shall discuss a particular definition for such a cumulative 
parameter, denoted
$\Delta_{\rm cu}$, which is purposefully chosen such that
$\Delta_{\rm cu}$ reduces to $\Delta$ in the bulk limit.  For small
grains, we show the correspondence of this definition to cumulative
properties such as the condensation energy.  We then define a 
single-particle parameter 
$\Delta_{\rm s.p.}$ and show the correspondence of this definition to
single-particle properties of the grain.  In the bulk limit both
$\Delta_{\rm cu}$ and $\Delta_{\rm s.p.}$ reduce to $\Delta$ as
expected, but at a different size scale (see Fig.~\ref{fig1}). 
For the single-particle
properties, we find $ \Delta_{\rm s.p.} \gg \Delta$ in the ultrasmall
regime ($d>\Delta$), and $ \Delta_{\rm s.p.} \simeq \Delta$ for larger
grains. In contrast, for the cumulative properties we find
$\Delta_{\rm cu} \gg \Delta$ not only in the ultrasmall regime, but
also in an ``intermediate regime''\cite{SILD01} in which
$\Delta>d>\Delta^2/\omega_{\rm D}$.  Regarding the
relation between cumulative and single-particle parameters, we find
$\Delta_{\rm cu} \gg \Delta_{\rm s.p.}$ in both the ultrasmall and
the intermediate regimes. 
Note that both the range of the intermediate regime, and the value of
the pairing parameters and the various superconducting properties
obtained below, depend explicitly on $\omega_{\rm D}$. This
$\omega_{\rm D}$-dependence, as well as the very existence of the
intermediate regime, are direct manifestations of the contribution of
the pairing correlations of the far levels.  Finally, we discuss some
possible strategies for measuring the single-particle and cumulative
quantities in small grains.  Throughout the paper, except when
discussing the even-odd effect, we assume for simplicity an even
number of electrons in the grain.

\section{Pairing correlations in the grand-canonical ensemble}
\label{sectwo}

We consider the reduced BCS Hamiltonian 
\begin{equation}
\hat{H} = \sum_{j,\sigma=\pm} \epsilon_j c^\dagger_{j\sigma} c_{j\sigma} 
- \lambda d \sum_{i,j} \, \hspace{-0.1cm} ' \, 
c^\dagger_{i+} c^\dagger_{i-} c_{j-} c_{j+} \; .
\label{Hamiltonian}
\end{equation} 
where the second sum (the pairing interaction) 
is restricted to energies within $\omega_{\rm D}$ of 
$E_{\rm F}$, and $+(-)$ denote spin up (down). 
The Hamiltonian~(\ref{Hamiltonian}) is the usual BCS Hamiltonian
used when discussing superconducting grains~\cite{DR01} 
and its validity is discussed in, e.g., Refs.~\onlinecite{AA97,Aga99,DR01}. 
For a grain with a given, finite number of electrons 
this Hamiltonian has an exact solution, obtained by 
Richardson\cite{Ric63,RS64}, and independently 
by Gaudin\cite{Gaud95}. 
In the macroscopic limit, where the canonical 
and grand-canonical ensembles produce the same results, Richardson's solution 
reduces to the BCS solution\cite{Ric77,RSD02}. 
As a result of the pairing interaction, the ground state of a superconductor
is different from the noninteracting Fermi state. It is a coherent 
superposition of various paired many-body noninteracting states, 
defined as Slater determinant of real one electron wavefunctions. 
The coherence means that the amplitudes for all these 
states in the superposition are real, up to an overall, global 
phase factor. This
is true both for the BCS wave function in the grand canonical
formalism

\begin{equation}
|{\rm BCS} \rangle = \prod_j (u_j + v_j b^\dagger_j)|{\rm Vac} \rangle 
\label{BCSwf}
\end{equation}  
where $b_j \equiv c_{j -} c_{j +}$, and for the exact wave
function given by Richardson's solution\cite{RS64}.  (In the grand
canonical formalism, the above coherence relates to noninteracting 
many-body states 
with a given number of pairs. One can add a constant phase between 
states of
different number of pairs, which is referred to as the superconducting
phase). A particular characteristic of the structure of the ground state 
of a superconductor is that the occupation probability of levels above the
Fermi energy is \emph{nonzero}, and that of levels below the Fermi
energy is smaller than unity.  A measure of the pairing correlations,
which exploits both the non Fermi like occupation probability and the
phase coherence mentioned above, is given by the pair amplitude, 
which in the g.c. ensemble is given by

\begin{equation}
\Delta_{\rm g.c.} \equiv \lambda d \sum_j \langle c_{j-} c_{j+} \rangle .
\label{deltagc}
\end{equation} 
For any  many-body BCS-like eigenstate, characterized by the set
$\{f_{j \sigma}\}$ of the occupation probabilities of the BCS
quasiparticles, one obtains\cite{Gen89} 

\begin{equation}
\Delta_{\rm g.c.}(\{f_{j \sigma}\})= \lambda d \sum_{j} u_j v_j^* 
(1-f_{j+}-f_{j-}) \; .
\label{bulklimitdelta}
\end{equation}
Specifically, for the ground state given in Eq.~(\ref{BCSwf}),
for which $f_{j \sigma} = 0$ for all $j$, we obtain 

\begin{equation}
\Delta_{\rm g.c.}({\rm g.s.}) = \lambda d \sum_j u_j v_j \, , 
\label{gcujvj}
\end{equation} 
which gives in the bulk limit $\Delta_{\rm g.c.}({\rm g.s.}) = \omega_{\rm
D}/\sinh(1/\lambda) \equiv \Delta$. The mean occupation of level $j$
is given by $v_j^2$, and $u_j^2 + v_j^2 = 1$. Note that the nature of
$\Delta_{\rm g.c.}$ is cumulative, being a sum of the contributions of
all levels. As a result of the phase coherence mentioned above, all the 
contributions to the sum in Eq.~(\ref{gcujvj}) come with the same phase.

\vspace{0.4cm}

\setlength{\unitlength}{3in}

\begin{figure} 
	\begin{center}  
	\begin{picture}(1,0.618)
	\put(-0.495,-2.45){\psfig{figure=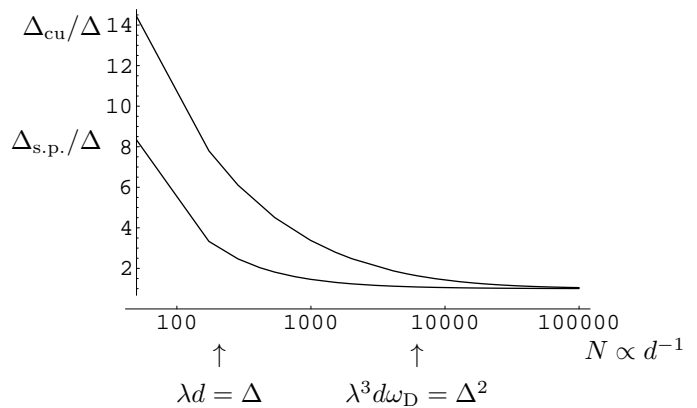,width=8.5in}}
	\put(1.0,0.03){\makebox(0,0)[t]{{$N \propto d^{-1}$}}}
        \put(0.275,0.02){\makebox(0,0)[t]{{$\uparrow$}}}
        \put(0.275,-0.05){\makebox(0,0)[t]{{$\lambda d = \Delta$}}}
        \put(0.62,0.02){\makebox(0,0)[t]{{$\uparrow$}}}
        \put(0.62,-0.045){\makebox(0,0)[t]{{$\lambda^3 d \omega_{\rm D} = \Delta^2$}}}
        \put(0.07,0.37){\makebox(0,0)[r]{{$\Delta_{\rm s.p.}/\Delta$}}}
        \put(0.07,0.57){\makebox(0,0)[r]{{$\Delta_{\rm cu}/\Delta$}}}
	\end{picture} 
	\end{center}
	\label{fig1}
        \vspace{0.8cm}  
        \caption{ The two pairing parameters, $\Delta_{\rm cu}$ (top curve) 
        and $\Delta_{\rm s.p.}$ (bottom curve), normalized to the value of 
        the bulk gap, are 
        sketched as function of $N=\omega_{\rm D}/d$ (proportional to the 
        inverse grain size), for $\lambda=0.12$. 
        $\Delta_{\rm s.p.} \gg \Delta$ for $N<250$, which correspond to 
        $\lambda d = \Delta$. $\Delta_{\rm cu} > \Delta, \Delta{\rm s.p.}$ 
        both in the ultrasmall regime, and in the intermediate 
          regime, up to $N \simeq 7500$ which corresponds to 
        $\lambda^3 d \omega_{\rm D} = \Delta^2$. 
	} 
        \vspace{0.5 cm}
\end{figure}

\section{Cumulative pairing parameter $\Delta_{\rm \lowercase{cu}}$}

We now turn to the definition of the cumulative canonical pairing
parameter. A natural extension of the concept of long range order in the bulk, 
suggests the following definition for a canonical pairing parameter: 

\begin{equation}
|\Delta_{\rm cu}|^2 \equiv (\lambda d)^2 \int d r_1 d r_2 F(r_1,r_2) 
\label{deltacdef}
\end{equation}
where $F(r_1,r_2)$ is a function characterizing pairing correlations, 
which is given by

\begin{eqnarray}
F(r_1,r_2) \equiv 
\langle \psi^\dagger_+(r_1) \psi^\dagger_-(r_1) 
\psi_-(r_2) \psi_+(r_2) \rangle  - \\ 
 \langle \psi^\dagger_+(r_1) \psi_+(r_2) 
\rangle \langle \psi^\dagger_-(r_1) \psi_-(r_2) \rangle \, . \nonumber
\label{fr1r2}
\end{eqnarray}
We will show that this definition for the cumulative canonical pairing 
parameter is adequate, as it has the following properties:  
(i) it is meaningful for a canonical ensemble; 
(ii) in the bulk limit 
it is equivalent to the grand canonical definition (\ref{deltagc}) 
{\it for any given BCS eigenstate;} 
(iii) it is related to the condensation energy and to the mean occupation 
of the noninteracting levels not only in the bulk limit, 
but also in the opposite limit of ultrasmall grains 
[see Eqs.~(\ref{condensationdelta}--\ref{deltaoccupation}) below].

Expanding each of the $\psi$ operators in the basis of the noninteracting 
single-particle eigenstates, we obtain 

\begin{equation} 
|\Delta_{\rm cu}|^2 = (\lambda d)^2 \sum_{ij} (\langle b^\dagger_i b_j 
\rangle - \langle c^\dagger_{i+} c_{j+} \rangle  \langle 
c^\dagger_{i-} c_{j-} \rangle) \, . 
\label{deltacano}
\end{equation} 
Since the terms in each of the brackets are number conserving, they
are meaningful in the canonical ensemble, which is used below for the
evaluation of $|\Delta_{\rm cu}|^2$ for small grains.  However, in
the macroscopic limit, one can use the grand canonical ensemble to
evaluate $|\Delta_{\rm cu}|^2$ within the BCS approximation.  By
using the Bogoliubov transformation and the BCS wavefunctions, we find
that for any many-body BCS-like eigenstate 

\begin{eqnarray}
|\Delta_{\rm cu}|^2 = & & (\lambda d)^2 \sum_{ij} u_i v_i^* u_j^* v_j 
(1-f_{i+}-f_{i-}) (1-f_{j+}-f_{j-}) \nonumber \\ 
= |\Delta_{\rm g.c.}|^2 , & & 
\label{bulklimit}
\end{eqnarray}
where the last equality is a result of Eq.~(\ref{bulklimitdelta}). 
The canonical pairing parameter is equal, in the bulk limit, to the
grand canonical pairing parameter for any many-body state, and
therefore both definitions are equivalent in this limit. 
Our definition for $\Delta_{\rm cu}$ differs slightly
from that given in Ref.~\onlinecite{RCR99}. The difference lies in the last
term in Eq.~(\ref{deltacano}), which results in the exact equivalence
to $\Delta_{\rm g.c.}$ in Eq.~(\ref{bulklimit}).

We now turn to the opposite limit of ultrasmall grains. By examining
Richardson's exact solution\cite{Ric63,RS64}, it was shown that
pairing correlations, however small [i.e.\ even for  $\lambda \ll
1/\ln{(\omega_{\rm D}/d)}$] manifest themselves both in the form of
the ground state wave function and in a finite condensation energy
$E_{\rm cond}$\cite{SILD01}. Here we make the connection between these
two effects and the canonical pairing parameter. In particular, we
shall show that to leading order in $\lambda$

\begin{equation}
E_{\rm cond} = \frac{|\Delta_{\rm cu}|^2}{2 \lambda d} ,
\label{condensationdelta}
\end{equation}
where $|\Delta_{\rm cu}|^2$ is evaluated for the exact ground state 
(this result differs from the known bulk result 
$E_{\rm cond} = \Delta^2/(2 d)$ by the occurrence of $\lambda$ in the 
denominator, which is discussed below). 
Also, $\Delta_{\rm cu}$ is
 related to the sum of the pairing correlations 
in all the levels, as are reflected in their occupation probabilities,
since we shall show that 

\begin{equation}
|\Delta_{\rm cu}|^2 = \frac{2 \ln{2}}{\lambda} \Delta_{\rm occ}^2 \,
 , 
\label{deltaoccupation}
\end{equation}
where 

\begin{equation}
\Delta_{\rm occ} \equiv \lambda d \sum_j \bar{u}_j \bar{v}_j \; , 
\label{deltaocc}
\end{equation} 
with $\bar{v}_j^2 \equiv \langle b^\dagger_j b_j \rangle$ and 
$\bar{u}_j^2 = 1 -  \bar{v}_j^2$.
The quantity $\Delta_{\rm occ}$, 
which has been defined in 
analogy to $\Delta_{\rm g.c.}$,
reflects the non-Fermi-like mean occupation probability in the many-body 
ground state of the noninteracting single-particle levels.

These results are found as follows:
To obtain the value of $|\Delta_{\rm cu}|^2$ in the ground state we
use Richardson's equations and expressions for the wavefunctions
\cite{RS64}, to write the ground state to leading order in $\lambda$,
in terms of the amplitudes of the various noninteracting many-body
states appearing in it:

\begin{eqnarray}
\label{gswf}
\phi_{\rm g.s.}(1,...,N) & \mbox{=} & 1 \nonumber \\ 
 \phi_{\rm g.s.}(1,...,N; \neq j, k) & \mbox{=} & 
\frac{\lambda d}{2(\epsilon_k - \epsilon_j)} \; .
\end{eqnarray} 
The first line refers to the amplitude that all the levels below the
Fermi energy are occupied by pairs (noninteracting ground state of the
system). The second line is the amplitude for the noninteracting
many-body state which is the same as the noninteracting ground state,
except for one pair excitation from level $j$ to level $k$. The
amplitudes of all the other many-body noninteracting states are zero
to first order in $\lambda$.

From Eq.~(\ref{gswf}), to zeroth order in $\lambda$, the two terms in 
Eq.~(\ref{deltacano}) cancel each other. 
The second term in Eq.~(\ref{deltacano}) has no
contribution to first order in $\lambda$. The contribution to the
first term which is first order in $\lambda$ comes from the fact that
there is finite amplitude for levels above the Fermi energy to be
occupied, and is given, using Eq.~(\ref{gswf}), by $\lambda
d/2(\epsilon_j - \epsilon_i)$ for each $j>,i<$ (by $<(>)$ we refer to
levels below (above) the Fermi energy). Since the sum in
Eq.~(\ref{deltacano}) is unrestricted, we get a factor of two, and to
leading order in $\lambda$

\begin{equation}
|\Delta_{\rm cu}|^2 = (\lambda d)^3  \sum_{i<,j>} \frac{1}{\epsilon_j - 
 \epsilon_i} \; .  
\label{deltacanfirstlambda}
\end{equation} 
The condensation energy was calculated in Ref.~\onlinecite{SILD01}. In order
to compare it with $\Delta_{\rm cu}$ of Eq.~(\ref{deltacanfirstlambda}) 
we present here its value for a general spectrum, which to leading
order in $\lambda$ is given by

\begin{equation} 
E_{\rm cond} = (\lambda d)^2 \sum_{i<,j>} \frac{1}{2 \epsilon_j - 
2 \epsilon_i} \, .
\label{condensationspectrum}
\end{equation} 
This result is obtained directly from Richardson's equations, and leads 
to Eq.~(\ref{condensationdelta}). 

Evaluating Eq.~(\ref{deltacanfirstlambda}) for equally-spaced 
spectrum, we obtain  

\begin{equation}
|\Delta_{\rm cu}|^2 = 2 \ln{2} \cdot \lambda ^3 \, d \; \omega_{\rm D} \; . 
\label{deltacandebye}
\end{equation} 
The large magnitude (linear in $\omega_{\rm D}$) 
of this result is due to the 
fact that all the amplitudes in (\ref{gswf}) have the same phase 
(which is a consequence of the coherence discussed in Sec.~\ref{sectwo}),
so that all the terms in Eq.~(\ref{deltacanfirstlambda}) are
added with the same sign.

In Ref.~\onlinecite{SILD01} we have shown that $E_{\rm cond} \simeq
E^{\rm BCS}_{\rm cond} + E^{\rm pert}_{\rm cond}$, where $E^{\rm
BCS}_{\rm cond}=\Delta^2/2d$ is the contribution of the condensed
levels, and $E^{\rm pert}_{\rm cond} \approx \lambda^2 \omega_{\rm D}$
is the contribution of the far (not-condensed) levels, which can be
calculated perturbatively. Similarly, $|\Delta_{\rm cu}|^2$ is a sum
of the contributions of all levels below $\omega_{\rm D}$, too, and is
related to the condensation energy in both the BCS and the
perturbative regimes. Therefore, it is natural to hypothesize that
$|\Delta_{\rm cu}|^2$ (similarly to $E_{\rm cond}$) can to a good
approximation be written as the sum of a bulk contribution from the
condensed levels and a perturbative contribution from the far levels,
i.e.\ $|\Delta_{\rm cu}|^2 \simeq 2 \ln 2 \lambda^3 d \omega_{\rm D} +
\Delta^2$.  This would imply that $\Delta_{\rm cu} \gg \Delta$ in both
the ultrasmall and intermediate regimes (see Fig~\ref{fig1}).  Note,
that the expression for the condensation energy in
Eq.~(\ref{condensationdelta}) is different from the bulk expression of
the condensation energy in terms of $\Delta$, namely $E^{\rm
bulk}_{\rm cond} = \Delta^2 / 2d$, by a factor of $\lambda$ in the
denominator. Since we have shown that in the bulk limit $\Delta_{\rm
cu}=\Delta_{\rm g.c.}$, we conclude that $\Delta_{\rm cu}$ and the
condensation energy have a different functional dependence on
$\lambda$. The above hypothesis results in

\begin{equation}
E_{\rm cond} = \frac{|\Delta_{\rm cu}|^2}{2 f(\lambda) d} \; , 
\label{continuouslambda}
\end{equation} 
where $f(\lambda)$ is a monotonic function of $\lambda$, which equals
$\lambda$ for $\lambda \ll 1/\ln{N}$ and unity for $\lambda \gg
1/\ln{N}$ (bulk limit). 

We now turn to derive the relation (\ref{deltaoccupation}) between
$\Delta_{\rm cu}$ and the mean occupation probabilities. In the exact
ground state, $\bar{v}_j^2$ is given by

\begin{equation} 
\sum_{\{f_1...f_N \mid j \in \}} |\phi(f_1,...,j,...f_N)|^2  \; , 
\label{f1fN}
\end{equation} 
where $\{f_1...f_N \mid j \in \}$ is any configuration of $N$ levels out 
of the $2 N$ noninteracting single-particle levels, 
that includes the level $j$. 
To leading order in $\lambda$, using Eq.~(\ref{gswf}), we obtain for
levels $j$ above $E_{\rm F}$

\begin{equation}
\bar{v}_j^2 = (\lambda d)^2 \sum_{i<} 
\frac{1}{4(\epsilon_j-\epsilon_i)^2} \, . 
\label{vjexplicit}
\end{equation}
Using the approximation of constant level spacing, we find that 

\begin{equation}
\bar{v}_j^2 = \frac{\lambda^2 d}{4 \epsilon_j} \; , 
\label{vjconstls}
\end{equation} 
where $\epsilon_j$ is measured from the Fermi energy.  
This important result shows that the 
mean occupation $\bar{v}^2(\epsilon)$ 
is proportional to $\epsilon^{-1}$, unlike the usual
BCS result, where for $\epsilon \gg \Delta$ the mean occupation is
proportional to $\epsilon^{-2}$. Since the occupation probability for
a single level is, by (\ref{vjconstls}), proportional to the level
spacing, this term can be neglected in the bulk limit. However, in
finite-size grains, we find that
in both the ultrasmall 
{\it and intermediate} regimes (i.e.\ 
$\Delta < \lambda \sqrt{d \omega_{\rm D}}$),
the occupation probability
for energies of the order of $\omega_{\rm D}$
is in fact larger than that given by the BCS
approximation.

As a result of Eq.~(\ref{vjexplicit}), to first order in $\lambda$ we 
find that 
$\bar{v}_j = \lambda d \sqrt{\sum_{i<} 1/4(\epsilon_j-\epsilon_i)^2}$ 
and $\bar{u}_j = 1$. 
For equally spaced levels, one therefore obtains 
from Eq.~(\ref{vjconstls}) that 

\begin{equation}
\Delta_{\rm occ} = \lambda d \sum_j \bar{u}_j \bar{v}_j = \lambda^2 
\sqrt{d \omega_{\rm D}} \; ,
\label{barujvj}
\end{equation} 
which yields the 
relation between $\Delta_{\rm occ}$ and $\Delta_{\rm cu}$ given in 
Eq.~(\ref{deltaoccupation}).  To summarize: the condensation 
energy, pairing parameter, and mean level occupation all 
receive significant contributions from the weak pairing 
correlation of the far (``not-condensed'') levels up to the Debye 
frequency. As a result, their magnitude is much larger than that 
given by the BCS approximation, not only in the ultrasmall regime, 
but also in the intermediate regime where 
$\Delta^2/\omega_{\rm D} < d < \Delta$.

\section{Single-particle pairing parameter $\Delta_{\rm \lowercase{s.p.}}$}

Unlike the cumulative properties considered above, other
superconducting properties, such as the excitation energy and the
even-odd effect (quantified by the Matveev--Larkin (ML) parameter, see
Ref.~\onlinecite{ML97}), are related to the blocking of, say, only one or
two levels to pairing correlations. Therefore, they do not depend on
the correlations between all possible pairs of levels in the grain,
but only on the correlations between these selected levels and all the
other levels. As a result, \emph{their values in small grains are much
smaller than one would get by using the bulk analogy with $\Delta_{\rm
cu}$ as the pairing parameter}.  We therefore define

\begin{equation}
\Delta_{\rm s.p.}^i \equiv \lambda d 
\sum_j ( \langle b^\dagger_i b_j \rangle + 
\langle b_i b^\dagger_j \rangle )  
\label{deltaspidef}
\end{equation}
where the sum is over the noninteracting single-particle levels and
$i$ is a selected such level. Since one frequently deals with the
lowest energy levels, we define

\begin{equation}
\Delta_{\rm s.p.} \equiv \Delta_{\rm s.p.}^{\bar{i}} 
\label{deltaspdef}
\end{equation}
for $\bar{i}$ being the level closest to the Fermi energy (for our
considerations the cases that $\bar{i}$ is below or above $E_{\rm
F}$ are equivalent, and we take $\bar{i}$ to be below $E_{\rm F}$).

In the bulk limit, using the BCS approximation, we obtain for the
ground state

\begin{equation}
\Delta_{\rm s.p.}^i = 2  \lambda d u_i v_i \sum_j u_j v_j \; , 
\label{deltaspiBCS}
\end{equation}
and specifically 
\begin{equation}
\Delta_{\rm s.p.} = \lambda d \sum_j u_j v_j = \Delta.
\label{deltaspBCS}
\end{equation} 
We now turn to the ultrasmall regime, and evaluate $\Delta_{\rm s.p.}$
in the ground state to second order in $\lambda$. Using
Eq.~(\ref{gswf}) we obtain 

\begin{equation}
\Delta_{\rm s.p.}^i = 
\lambda d + 
\sum_{j>} \frac{(\lambda d)^2}{\epsilon_j - \epsilon_i} \; ,
\label{deltaspgs}
\end{equation}
and for the equally spaced spectrum 

\begin{equation}
\Delta_{\rm s.p.} =  
\lambda d +  
\lambda^2 d \ln{\frac{\omega_{\rm D}}{d}} \; .
\label{deltaspgses}
\end{equation}

Let us now consider the excitation energy, say $E_{\rm exc}$, 
of the first excited 
state of an (even) superconducting grain. This state can be described
by having the two levels nearest to $E_{\rm F}$ singly occupied with
probability unity, thus breaking one pair, and the other $N -1$ pairs
occupying the remaining $2 N -2$ levels according to Richardson's
exact solution \cite{RS64}.  The excitation energy has two different
contributions.  (i) The kinetic energy cost $d$ of occupying a level of
higher energy with probability unity, and (ii) a pairing energy
cost, which is given by $\Delta_{\rm s.p.}$. Note that the latter
has (iia) a diagonal part, given by $\lambda d$, which is
related to the excess energy of two electrons occupying the same
spatial noninteracting eigenstate, and (iib) an off-diagonal part due
to the blocking of the two singly occupied levels to pairing
correlations. Adding these contributions gives

\begin{equation}
E_{\rm exc} = d + \Delta_{\rm s.p.} \approx 
d + \lambda d +  \lambda^2 d \ln{\frac{\omega_{\rm D}}{d}} \; ,
\label{fexc}
\end{equation}
where the last expression is valid for a grain with equally spaced
spectrum. Similarly, the ML parameter\cite{ML97} is given by

\begin{equation}
\Delta_{ML} = \frac{1}{2} \Delta_{\rm s.p.} \; . 
\label{MLsp}
\end{equation}
Note that the pairing contribution to the
energy cost of the first excitation equals $2
\Delta_{\rm s.p.}$ in the bulk limit, whereas it 
equals $\Delta_{\rm s.p.}$ in an
ultrasmall grain, and the same factor of two appears in the ML
parameter. This reflects the fact that
in the bulk limit each level is correlated with
all the other levels, while in the ultrasmall grain, the dominant
correlations are between levels on different sides of the Fermi
energy.

\section{Measurement strategies for $\Delta_{\rm \lowercase{cu}}$ and 
$\Delta_{\rm \lowercase{s.p.}}$} 

A possible measurement of the cumulative
correlations of a superconducting grain was discussed in
Ref.~\onlinecite{SILD01}. It was shown that the condensation energy can be
obtained from specific heat or spin magnetization measurements, where
an explicit calculation was done for the latter.

In distinction with the case of bulk superconductivity, where the 
single-particle
properties are easiest to measure through the lowest excitations of
the system, in small grains the energy of the first excited state is
given by Eq.~(\ref{fexc}), in which $\Delta_{\rm s.p.}$ is only a
small correction to the level spacing.  Matveev and Larkin suggested
to measure their pairing parameter by the parity-induced alternation
of Coulomb blockade peak spacings in small grains. Here we suggest the
possibility that for an ensemble of small, weakly-coupled grains,
$\Delta_{\rm s.p.}$ could be measured through their spin magnetization
as a function of magnetic field at zero temperature.  
We assume that the coupling between the grains is weak enough such that 
the equilibrium properties of the system can be approximated by summing over 
the individual grains, but strong enough so tunneling between grains in 
a large enough ensemble (of say 50 grains) occurs within the time of 
the experiment. 

For a single-grain, the spin magnetization shows steps at values that
correspond to the Zeeman energy being equal the energy required 
to break a pair, of which the first is  at $E_{\rm exc} = d +
\Delta_{\rm s.p.}$ [by Eq.~(\ref{fexc})], which is dominated by the
large \emph{single-grain} level spacing $d$. In order to avoid having
to worry about the latter, we consider an ensemble of weakly connected
superconducting grains, which would have an effective joint level
spacing $d_{\rm ens}$ that is much smaller than $d$ if the ensemble is
large, $d_{\rm ens} \ll d$ (we neglect the charging energy. Note, that 
the charging energy was found to be much smaller\cite{ASCA75,MS79}
than what naive estimates give. Since the tunneling 
between the grains is weak, the relevant ensemble is the canonical one). 
Then the ground state of
the unconnected system is given by each of the grains having an even
number of electrons, being in its ground state.  
The first excited state of the system 
is given by moving one electron between
two grains, so that afterwards 
the first has a single-occupied level just
below $E_{\rm F}$, and the second has a singly-occupied level just
above $E_{\rm F}$. For a system of ``normal'' noninteracting grains
the energy of such an excitation is of the order of the single-particle 
level spacing
of the whole ensemble of grains, $d_{\rm ens}$. However, in a system
of ultrasmall superconducting grains, since two singly occupied levels
are created in two different grains, the energy of such an excitation
would be given by $\Delta_{\rm s.p.}$ (which is $\gg d_{\rm ens}$).
Since the first step in the spin magnetization is at the energy of the
first excited state of the system, in such a measurement we predict a
gap of value $\Delta_{\rm s.p.}$. (In a similar measurement in grains
with $d < \Delta$, the measured gap would be given by $\Delta$). 
Thus,  $\Delta_{\rm s.p.}$ can be interpreted as 
the superconducting gap as measured by single particle
properties. 
In principle, performing such a measurement as a function of the grain
size would monitor the change of $\Delta_{\rm s.p.}$, from being much
larger than $\Delta$ for ultrasmall grains to equaling $\Delta$ for
large ones. Note by that reducing the grain size one {\it increases}
this gap $\Delta_{\rm s.p.}$. 

Finally, we would like to note that the logarithmic dependence of the
correlation (second) term of $\Delta_{\rm s.p.}$ in
Eq.~(\ref{deltaspgses}) is manifested in the interaction correction to
the ensemble averaged magnetic response of small metallic grains, when 
considered within the BCS model\cite{SOIL02}.

\section{Summary}

We have shown that various superconducting properties can be
classified to two groups, one containing those properties which are
single particle in nature, and the other containing those properties
which are cumulative, a result of the summed contributions of many levels. 
Unlike the case in bulk
superconductivity, where both these properties are characterized by
one energy parameter $\Delta$, in general, and in particular in small
grains, two different energy parameters, $\Delta_{\rm s.p.}$ for the
former group and $\Delta_{\rm cu}$ for the latter one, characterize
superconducting properties. 

In bulk superconductivity the relevant contribution to the various 
superconducting properties comes from the ``condensed'' levels, within 
$\Delta$ of $E_{\rm F}$. However, in small grains the contribution 
of all levels up to $\omega_{\rm D}$ is significant, and this results 
in the existence of these two different parameters, as well as in 
the fact that 
$\Delta_{\rm cu} \gg \Delta_{\rm s.p.} \gg \Delta$ in the ultrasmall 
regime and $\Delta_{\rm cu} \gg \Delta$ also in the intermediate regime. 
Experimental possibilities to measure the two pairing parameters were 
discussed. 

\acknowledgments

M.S. is thankful for the support by the Lady Davis fund. 
This work was supported by a Center of Excellence of the Israel
Science Foundation, Jerusalem and by the German Federal Ministry of
Education and Research (BMBF) within the Framework of the
German-Israeli Project Cooperation (DIP) and by the German-Israeli
Foundation(GIF).

\end{multicols}

\end{document}